\documentclass[prl,twocolumn,showpacs]{revtex4}
\pdfoutput = 1

\usepackage{graphicx}
\usepackage{dcolumn}
\usepackage{bm}
\usepackage{amsmath}
\usepackage{amssymb}
\usepackage{mathrsfs}
\usepackage{amsfonts}
\usepackage{color}

\begin{document}

\title{Ancillary qubit spectroscopy of cavity (circuit) QED vacua}

\author{Jared Lolli$^1$, Alexandre Baksic$^{1}$, David Nagy$^1$, Vladimir E. Manucharyan$^2$, Cristiano Ciuti$^1$}
\affiliation{$^1$ Laboratoire Mat\'eriaux et Ph\'enom\`enes Quantiques,
Universit\'e Paris Diderot-Paris 7 and CNRS, \\ B\^atiment Condorcet, 10 rue
Alice Domon et L\'eonie Duquet, 75205 Paris Cedex 13, France}
\affiliation{$^2$ Department of Physics, University of Maryland, College Park, MD 20742.}

\begin{abstract}
We investigate theoretically how the spectroscopy of an ancillary qubit can probe cavity (circuit) QED  ground states containing photons. We consider three classes of  systems (Dicke, Tavis-Cummings and Hopfield-like models), where non-trivial vacua are the result of ultrastrong coupling between $N$ two-level systems and a single-mode bosonic field. An ancillary qubit detuned with respect to the boson frequency is shown to reveal distinct spectral signatures depending on the type of vacua. In particular, the Lamb shift of the ancilla is sensitive to both ground state photon population and correlations. Back-action of the ancilla on the cavity ground state is investigated, taking into account the dissipation via a consistent master equation for the ultrastrong coupling regime. The conditions for high-fidelity measurements are determined.
\end{abstract}
\maketitle

In recent years, cavity quantum electrodynamics (QED) has thrived thanks to the possibility of controlling light-matter interaction at the quantum level, which is relevant for the study of fundamental quantum phenomena, the generation of artificial quantum systems, and quantum information applications \cite{Haroche}. The field has more recently blossomed in solid-state systems, particularly in superconducting quantum circuit QED \cite{Wallraff,Blais2004} and semiconductor cavity QED \cite{Imamoglu}.
In conventional cavity QED, photons are present only in the excited light-matter states of the system and can escape the cavity due to a finite transparency of the mirrors. The situation changes drastically in the so-called ultrastrong light-matter coupling regime\cite{CiutiPRB2005,Devoret,Anappara,Todorov,Scalari,Gross}, achieved when the light-matter interaction rate is comparable or larger than the photon frequency. Indeed, it can become energetically
favorable to have photons in the ground state. However, such ground state photons are bound to the cavity and cannot escape, since that would violate the energy conservation\cite{CiutiPRA2006}.

In the 'thermodynamic' limit where a large number $N$ of two-level systems are (ultra)strongly coupled to a single bosonic mode, phase transitions can occur with non-trivial and degenerate vacua.
The vacuum properties depend on the details of the light-matter coupling and on the Hamiltonian symmetries. These phase transitions are associated with a symmetry breaking: it is a discrete $\mathbb{Z}_2$ symmetry for the phase transition \cite{Carmichael,Brandes,Dimer,esslinger} in the Dicke model \cite{Dicke} where  the non-rotating wave terms of light-matter interaction are considered; it is a continuous $U(1)$ symmetry in the case of the Tavis-Cummings model \cite{TavisCummings,Lieb} where non-rotating wave terms are absent. Such symmetries can be controlled in models where the two-level systems are coupled to both the quadratures of the bosonic field \cite{BaksicPRL}. On the other hand, in Hamiltonians containing a strong enough quadratic renormalization of the cavity photon frequency (e.g., due to the squared electromagnetic vector potential term), the ground state is a two-mode squeezed vacuum, but no phase transition occurs \cite{NatafNAT}.
This is the case for the Hopfield model \cite{Hopfield}, notably realized in semiconductor microcavities \cite{CiutiPRB2005,Anappara,Todorov,Scalari}.
The fundamental meaning and validity of cavity and circuit QED quantization procedures is critically at play in the ultrastrong coupling regime, since different forms of Hamiltonians lead to extremely different physical phenomena \cite{Viehmann,CiutiComment,Domokos}. Protocols to detect the properties of cavity vacua are therefore of strong significance, not only for a study of intriguing ground states, but also as a sensitive test of fundamental microscopic theories.

In this context, a crucial question arises: how to detect ground state photon populations and correlations without destroying them? In principle bound photons in cavity (circuit) QED vacua can be released by a non-adiabatic, ultrafast modulation of the Hamiltonian parameters \cite{CiutiPRB2005,DeLiberatoPRL,NoriPRL,DeLiberatoPRA,Delsing}, which can convert a ground state into an excited state.  While non-adiabatic QED provides an interesting way of observing non-classical quantum vacuum radiation, finding a non-invasive and sensitive probe of the ground state properties remains highly desirable. A theoretical work \cite{Lambert} in circuit QED has suggested to study the coupling between a Dicke system and an additional superconducting qubit, showing that Dicke criticality can be observed via current transport measurements. However, in Ref.  \cite{Lambert} the considered effective dispersive interaction between the cavity system and the auxiliary qubit depends only on the cavity photon population, and not on the intracavity light-matter correlations; moreover the dissipation has not been treated with a master equation suited for the ultrastrong coupling regime \cite{Beaudoin}, which is essential to avoid artifacts such the instability of the ground state and the excitation of the system in the absence of driving\cite{CiutiPRA2006,Beaudoin}.

In this Letter, we show that the spectroscopy of an ancillary qubit, moderately coupled to a cavity QED system, depends sensitively on the type of vacua.
By driving this ancillary qubit and analyzing its frequency response, we show that it is possible to have distinct signatures of the ground state photon populations and correlations without destroying them. We explore this protocol by considering three different classes of systems described respectively by the Dicke, Tavis-Cummings and Hopfield models, each one having a ground state with different properties.  We show numerically and analytically that the Lamb shift of the ancillary qubit transition is very sensitive both on the photon populations and correlations of exotic vacua. We explore the back-action of the ancillary qubit on the cavity ground state and determine the key physical quantities affecting the fidelity of the measurement, including consistently the dissipation effects in the ultrastrong coupling regime.

\begin{figure}[t!]
 \begin{center}
    \includegraphics[width=160pt,angle=270]{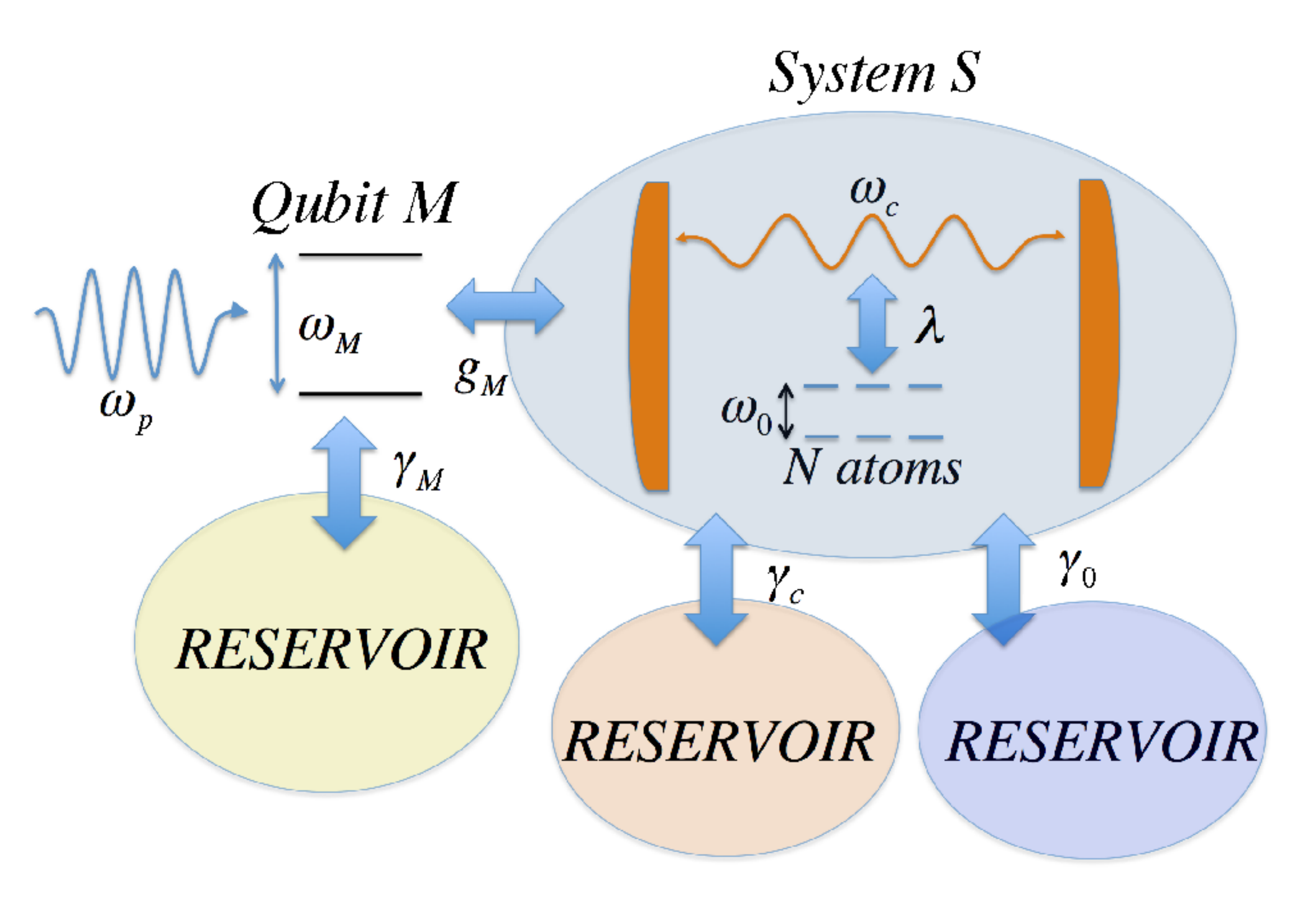}
   \end{center}
 \caption{
A sketch of a cavity (circuit) QED system $S$ consisting of a single-mode resonator coupled to $N$ two-level (artificial) atoms. An ancillary qubit $M$ is coupled to the resonator boson mode. The spectroscopy of the ancilla is used to probe quantum ground state properties of $S$.
\label{sketch}}
\end{figure}

\begin{figure}[h!]
   \begin{center}
   \includegraphics[width=220pt]{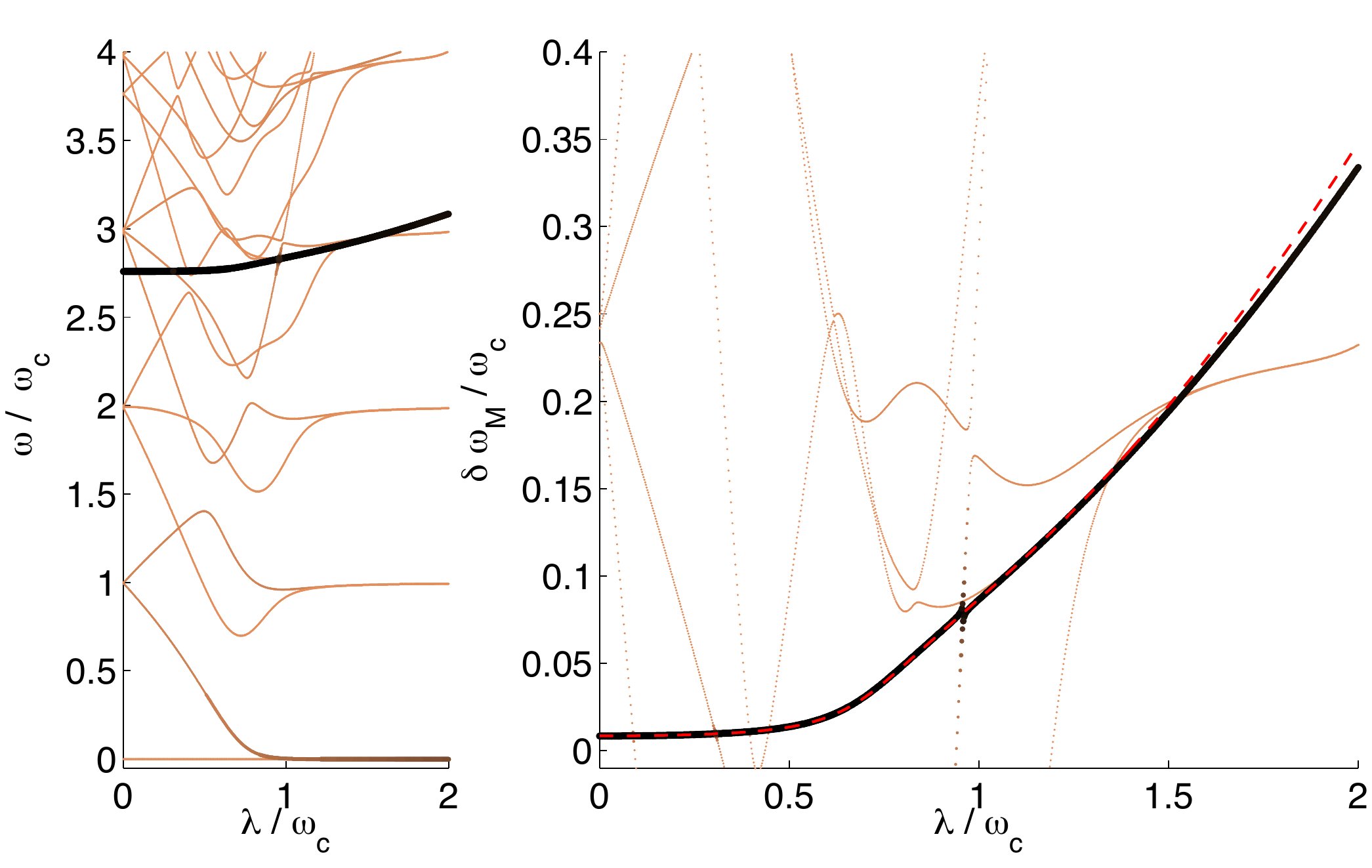}
   \\
   \includegraphics[width=220pt]{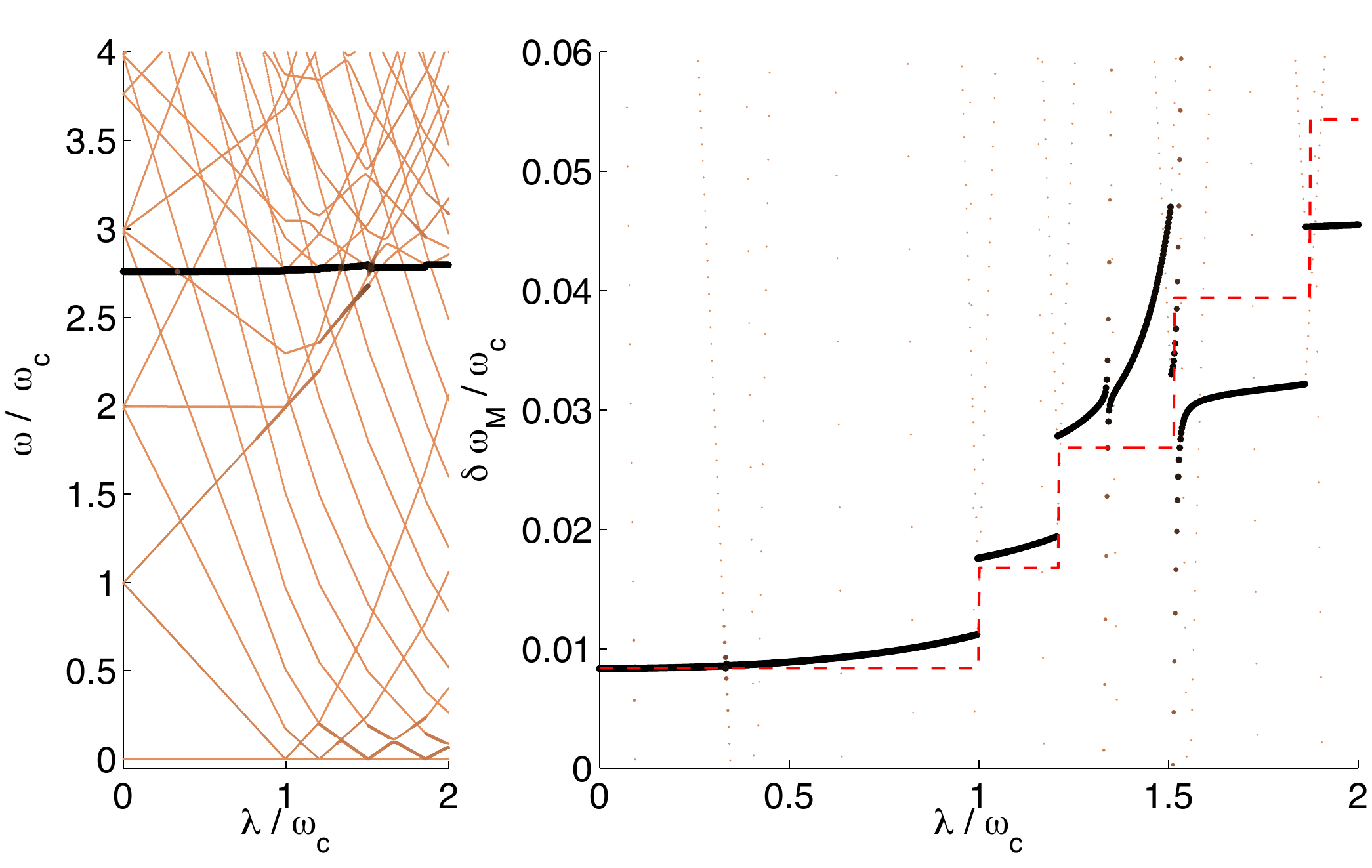}
   \\
     \includegraphics[width=220pt]{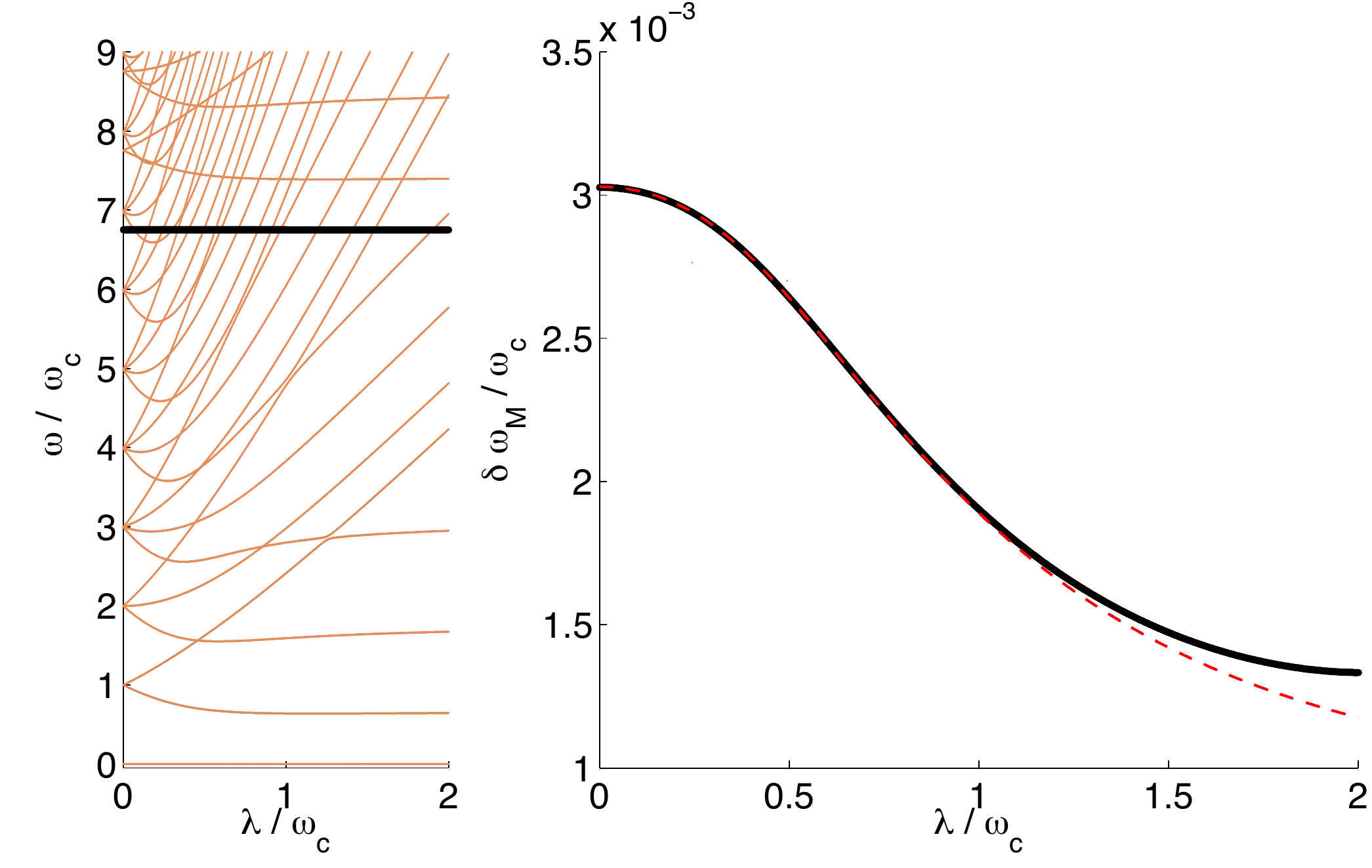}
   \end{center}
   \caption{(Color online)
 Left panels: excitation energies for the three considered systems $S$ versus the coupling $\lambda$ between the boson field and the $N$ atoms.
 Right panels: the Lamb shift of the ancillary qubit transition. The red dashed lines are calculated via Eq.~(\ref{dispersive}).
Top panels: Dicke system with $N=3$, $\omega_c=\omega_0$, $\omega_M=2.75 \omega_c$, $g_M=0.1 \omega_c$.
Middle panels: Tavis-Cummings system with $N=3$, $\omega_c= \omega_0$, $\omega_M=2.75 \omega_c$, $g_M=0.1 \omega_c$.
Bottom panels: an Hopfield system with $N=3$, $\omega_c= \omega_0$, $\omega_M=6.75 \omega_c$, $g_M=0.1 \omega_c$, $D = \lambda^2/\omega_0$.}
   \label{threemodels}
\end{figure}
\begin{figure}[t!]
 \begin{center}
   \includegraphics[width=121pt]{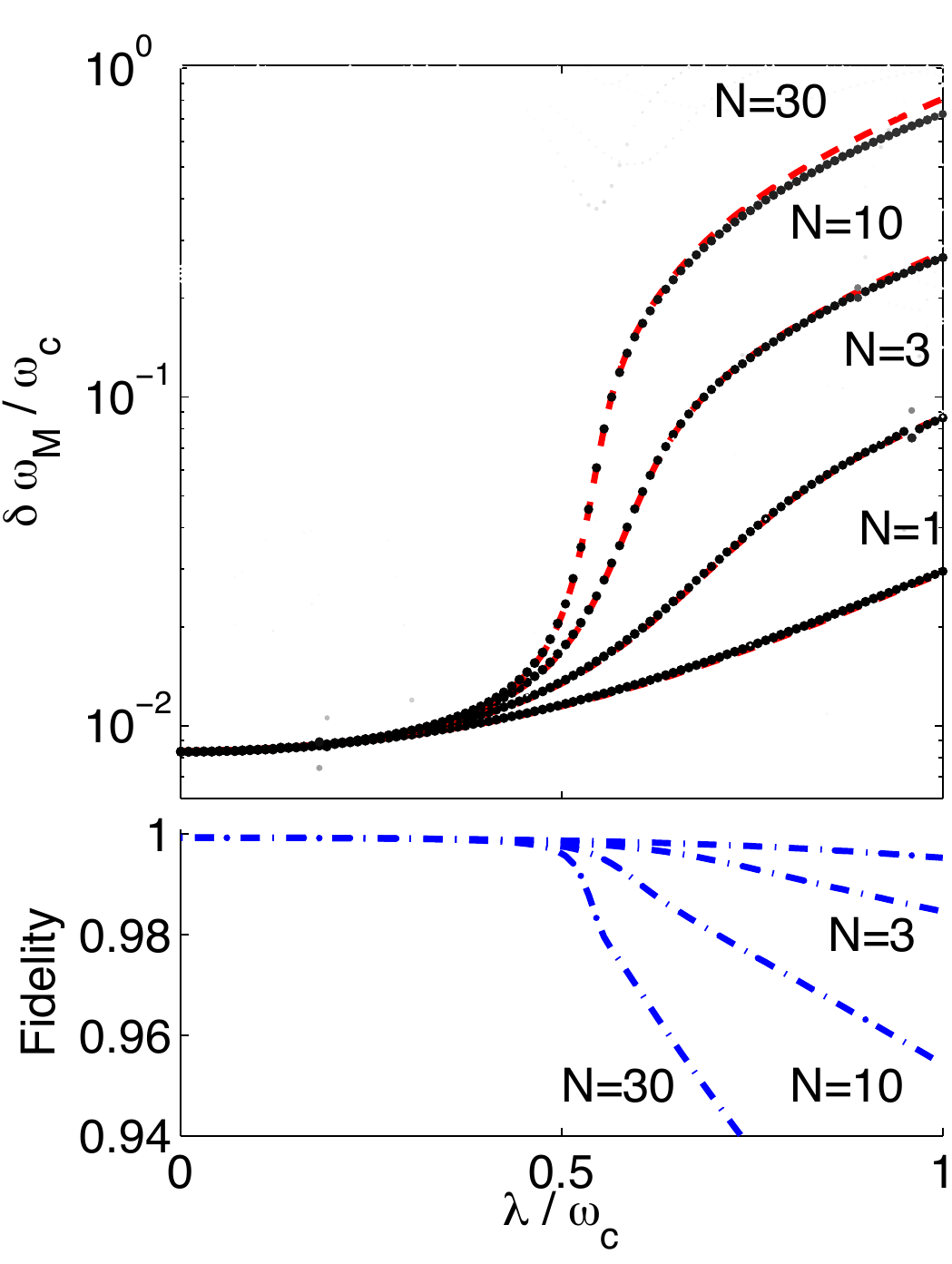}
   \includegraphics[width=121pt]{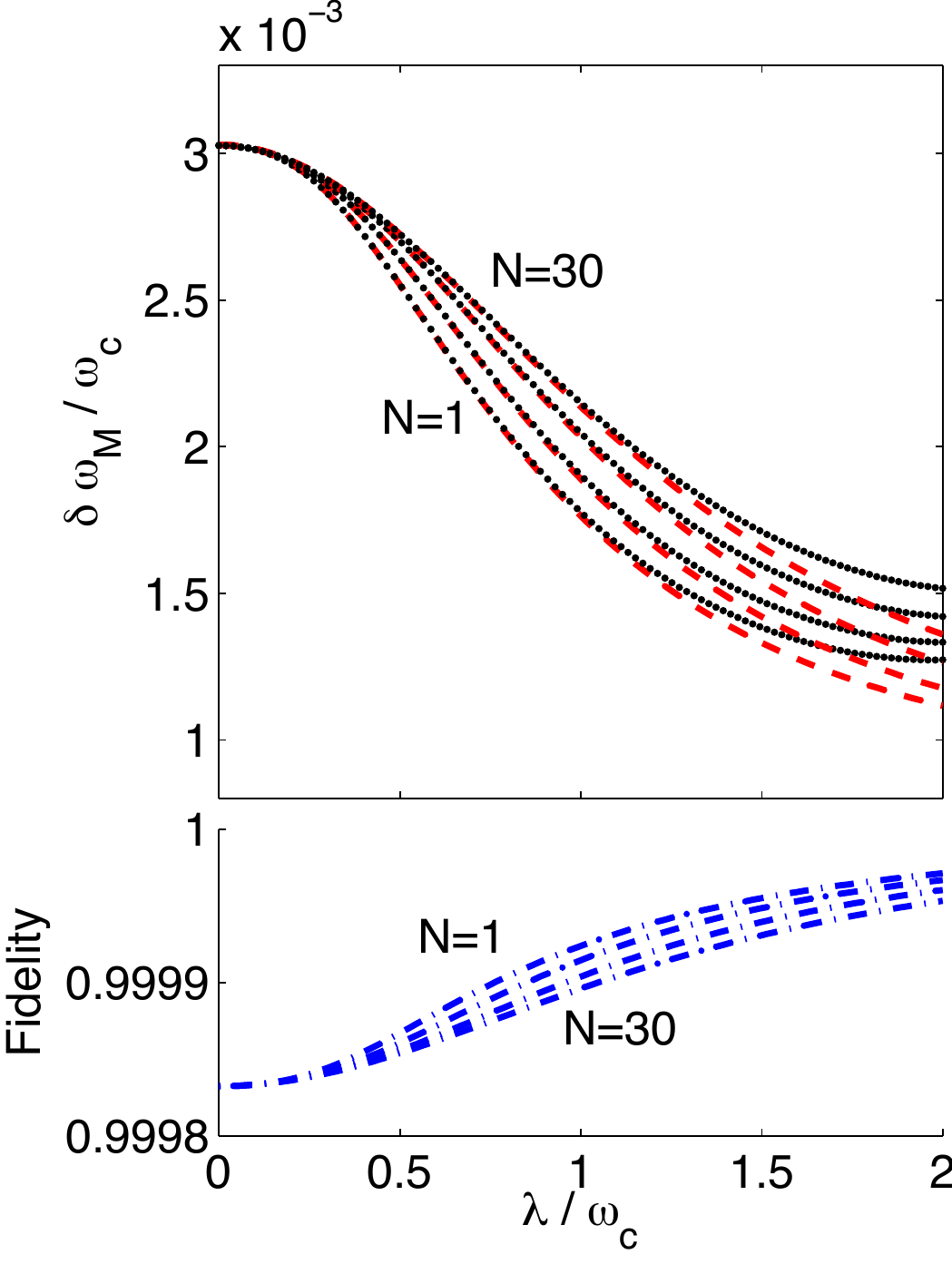}
   \end{center}
 \caption{
Top panels: Lamb shift of the ancillary qubit (black dots) versus $\lambda$ for $N=1,3,10 \text{ and } 30$. Red-dashed lines are obtained via Eq.~(\ref{dispersive}). Bottom panel: ground state fidelity $\mathcal{F}_G$.
Left panels: Dicke model with $\omega_c=\omega_0$, $\omega_M=2.75 \omega_c$, $g_M=0.1 \omega_c$.
Right panels:  Hopfield model with $\omega_c=\omega_0$, $\omega_M=6.75 \omega_c$, $g_M=0.1 \omega_c$, $D=\lambda^2/\omega_0$.
 \label{N_dependence}
}
\end{figure}

As sketched in Fig.~\ref{sketch}, we will consider an ancillary qubit $M$ coupled to the bosonic mode of a cavity (circuit) QED system $S$.
In particular, we will deal with the following time dependent Hamiltonian ($\hbar = 1$),
\begin{equation}
\mathcal{H}(t) = \mathcal{H}_S +\frac{\omega_M}{2} \sigma^{(M)}_z + g_M \left( a^{\dagger} + a \right) \sigma_x^{(M)} + \Omega_p \cos(\omega_p t)\sigma_x^{(M)} ,
\label{drivenH}
\end{equation}
where $\mathcal{H}_S$ is the system Hamiltonian, $g_M$ is the coupling between the measurement qubit and the boson field, whose boson annihilation operator is $a$.
The $\sigma^{(M)}_{i}$ Pauli operators act on the Hilbert space of the qubit $M$, whose transition frequency is $\omega_M$, while
$\Omega_p$ is the amplitude of the driving field with frequency $\omega_p$ acting on $M$.

In the following, $\mathcal{H}_S$ will be one of the three Hamiltonians, describing respectively the Dicke, Tavis-Cummings and Hopfield-like models ($\hbar=1$):
$$
\mathcal{H}_{Dicke} = \omega_c a^{\dagger}a
+\omega_0 J_z
+\frac{\lambda}{\sqrt{N}} \left( a^{\dagger} + a \right)\left( J_+ + J_- \right) ,
$$ $$
\mathcal{H}_{TC} = \omega_c a^{\dagger}a
+\omega_0 J_z
+\frac{\lambda}{\sqrt{N}} \left( a^{\dagger}J_- + a J_+\right) ,
$$ $$
\mathcal{H}_{Hopfield} = \mathcal{H}_{Dicke}  +D{\left( a^{\dagger} + a \right)}^2 ,
$$
where $\omega_c$ is the frequency of the bosonic mode, $\omega_0$ is the transition frequency of each of the $N$ identical two-level atoms, $\lambda$ is the collective coupling and $D = \lambda^2 / \omega_0$ is the strength of the boson renormalization term in the Hopfield model.
The $J_i$ are the angular momentum operators representing the collective pseudo-spin associated to the $N$ two-level systems, namely $J_z=\frac{1}{2}  \sum^N_{i=1} \sigma_z^{(i)}$, $J_{\pm}=\sum^N_{i=1} \sigma_{\pm}^{(i)}$, where the Pauli matrices here refer to each two-level system.

We start by considering the energy levels of $\mathcal{H}_{S+M}$,
describing the system $S$ coupled to the measurement qubit $M$, namely:
\begin{equation}
\mathcal{H}_{S+M}= \mathcal{H}_S +\frac{\omega_M}{2} \sigma^{(M)}_z + g_M \left( a^{\dagger} + a \right) \sigma_x^{(M)}.
\label{S+M}
\end{equation}
The eigenstates and their energies are defined by $\mathcal{H}_{S+M} \vert l \rangle = \epsilon_l \vert l \rangle$.
System $S$ will be of the Dicke, Tavis-Cummings or Hopfield type, as shown in Fig.~\ref{threemodels}.
We consider a qubit transition frequency $\omega_M$  detuned with respect to the boson frequency $\omega_c = \omega_0$.
For $g_M = 0$, the driving field term, proportional to the operator $\sigma_x^{(M)}$, induces a transition from the ground state $\vert G_S \rangle \otimes \vert \downarrow \rangle$ to the excited state $\vert G_S \rangle \otimes \vert \uparrow \rangle$, being
$\vert G_S \rangle$ the ground state of $S$ and
$\vert \downarrow \rangle$ ($ \vert \uparrow \rangle$) the ground (excited) state of the qubit $M$. For finite $g_M$, the coupling between $S$ and $M$ creates a mixing between states of the form $\vert \Psi_S \rangle \otimes \vert \psi_M\rangle$ and the driving induces a transition from the ground state $\vert G_{S+M} \rangle$ to excited states of $\mathcal{H}_{S+M}$.
Therefore, in the spectroscopy the relevant excited states $\vert l \rangle$ are those having the largest values of  $\vert \langle G_{S+M} \vert
\hat{\sigma}_x^{(M)}  \vert l \rangle \vert ^2$.
The color scale of the levels in Fig.~\ref{threemodels} is proportional to such matrix element. The results show that, due to the off-resonant coupling,  there is only one dominant spectroscopically-active level (black thick solid line), which has a strong overlap with the state $\vert G_S \rangle \otimes \vert \uparrow \rangle$.
  The right panels shows the Lamb shift of the qubit transition frequency. The top, middle and bottom panels of Fig.~\ref{threemodels} are respectively for the Dicke, Tavis-Cummings and Hopfield models.
  For $\lambda = 0$ the system $S$ is like a bare cavity, so the spectral renormalization is the standard Lamb shift \cite{Lamb_circuitQED} of the qubit due to the coupling to the normal vacuum in the cavity. By increasing $\lambda$ the vacuum is modified and the Lamb-shift changes.
   It is apparent that the behavior of the qubit shift is qualitatively different in the three cases.  For the Dicke model (top panels), the Lamb shift increases strongly with $\lambda$ and becomes much bigger than in the case of the bare cavity ($\lambda = 0$).
 In the Tavis-Cummings case (middle panels),  the Lamb shift increases in a step-like way as a function of  $\lambda$.  In the Hopfield model (bottom panels), instead the Lamb shift decreases with increasing value of $\lambda$.

By generalizing the approach in Ref. \cite{ZuecoPRA2009}, we have derived an analytical expression at the second order in $g_M$ for the measurement qubit Lamb shift, namely
\begin{widetext}
\begin{equation}
\delta \omega^{(S)}_M \simeq g_M^2  \left ( \frac{1}{ \omega_M-\omega_c} +\frac{1}{ \omega_M + \omega_c}  \right )  \langle (a+ a^{\dagger})^2  \rangle   +  g_M^2  \left ( \frac{1}{\left ( \omega_M-\omega_c \right )^2} -\frac{1}{ \left ( \omega_M + \omega_c \right )^2}  \right ) \langle \hat{V}^{(S)} \rangle  ,
\label{dispersive}
\end{equation}
\end{widetext}
where $\hat{V}^{(Dicke)} = \frac{\lambda}{\sqrt{N}} (a+ a^{\dagger})J_x$,  $\hat{V}^{(TC)} = \frac{\lambda}{\sqrt{N}} (a J_+ + a^{\dagger}J_-)$ and $\hat{V}^{(Hopfield)} = \hat{V}^{(Dicke)} + 2 \frac{\lambda^2}{\omega_0} (a+ a^{\dagger})^2 $.
Here the expectation values are calculated on the ground state $\vert G_S\rangle$ of the target system $S$.
Importantly, the shift not only depends on the ground state photon population $\langle a^{\dagger} a \rangle$, but also on the anomalous expectation value $\langle a^{\dagger  2} + a^2 \rangle$ and on the correlation between photon field and the $N$ two-level systems.
The red-dashed lines in the right panels of Fig.~\ref{threemodels} depict the shift predicted by Eq.~(\ref{dispersive}). The agreement between the numerical diagonalization results and the analytical formula is excellent in the considered range of values for $\lambda/\omega_c$, except for points where there are avoided crossings with other levels.

In Fig.~\ref{N_dependence}, we present the behavior of the qubit spectral shift as a function of $N$. With increasing value of $N$,  a critical point emerges for the Dicke Hamiltonian (left panels), but not for the Hopfield case (right panels). The behavior of the qubit Lamb shift, already completely different for small values of $N$, becomes strikingly dissimilar. Already for $N = 30$, the shift has a considerable jump around the critical coupling. The bottom panels shows the ground state fidelity $\mathcal{F}_G=Tr_{S,M}( \vert G_{S+M} \rangle\langle G_{S+M} \vert ( \vert G_S \rangle \langle G_S \vert \otimes \hat{1}^{(M)}) )$, quantifying the change of the cavity system ground state in presence of the ancilla qubit. In the considered conditions, $\mathcal{F}_G$ can be  close to $1$. However, for large values of $\lambda /\omega_c$ the fidelity strongly decreases in the Dicke case above the critical coupling, while it stays close to $1$ for Hopfield. Since for $g_M \to 0 $  the fidelity tends to $1$ and the qubit shift is proportional to $g_M^2$, a trade-off between fidelity and size of the qubit shift can be found depending on the degree of level broadening.
\begin{figure}[t!]
 \begin{center}
    \includegraphics[width=120pt]{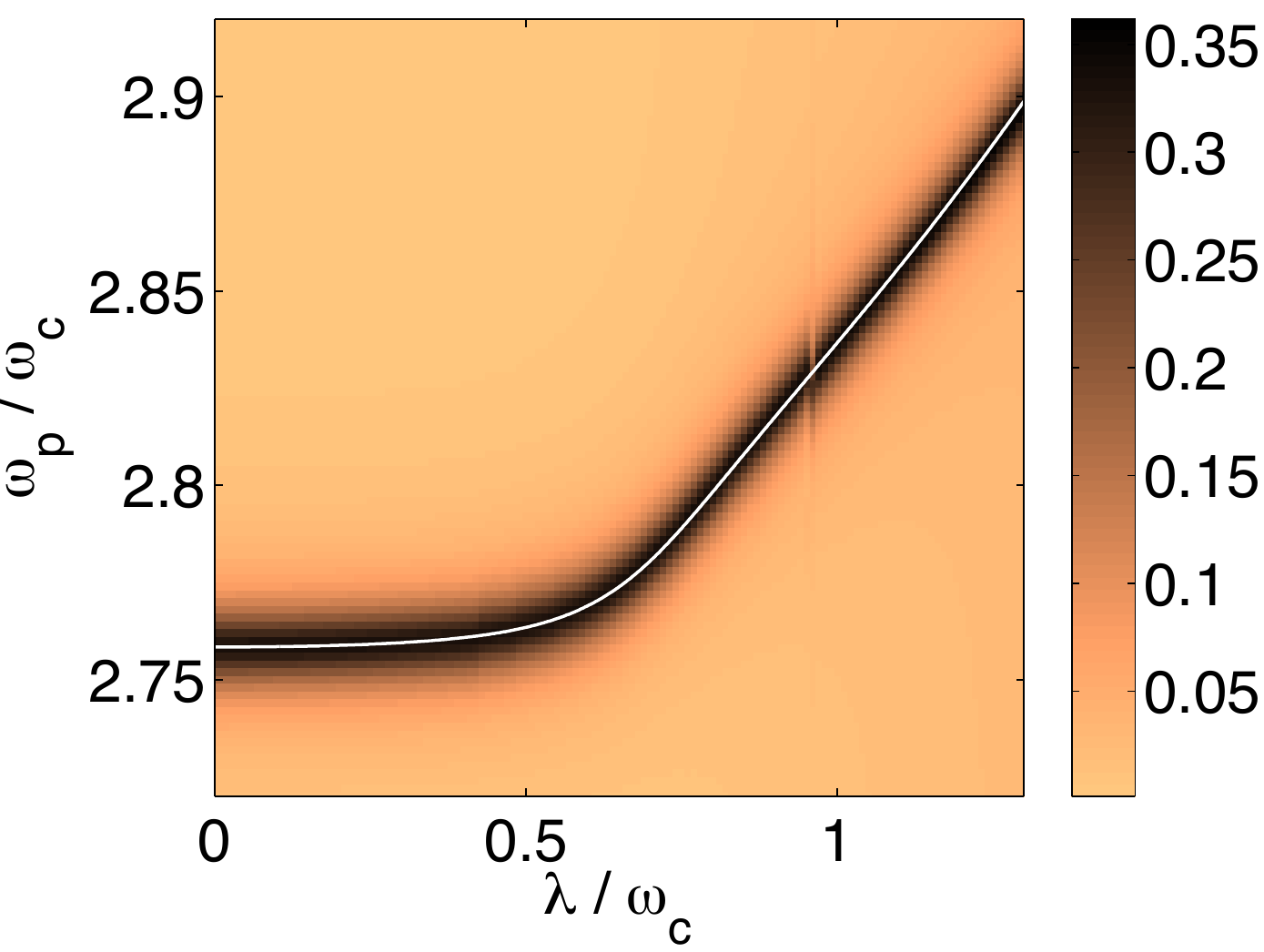}  \includegraphics[width=120pt]{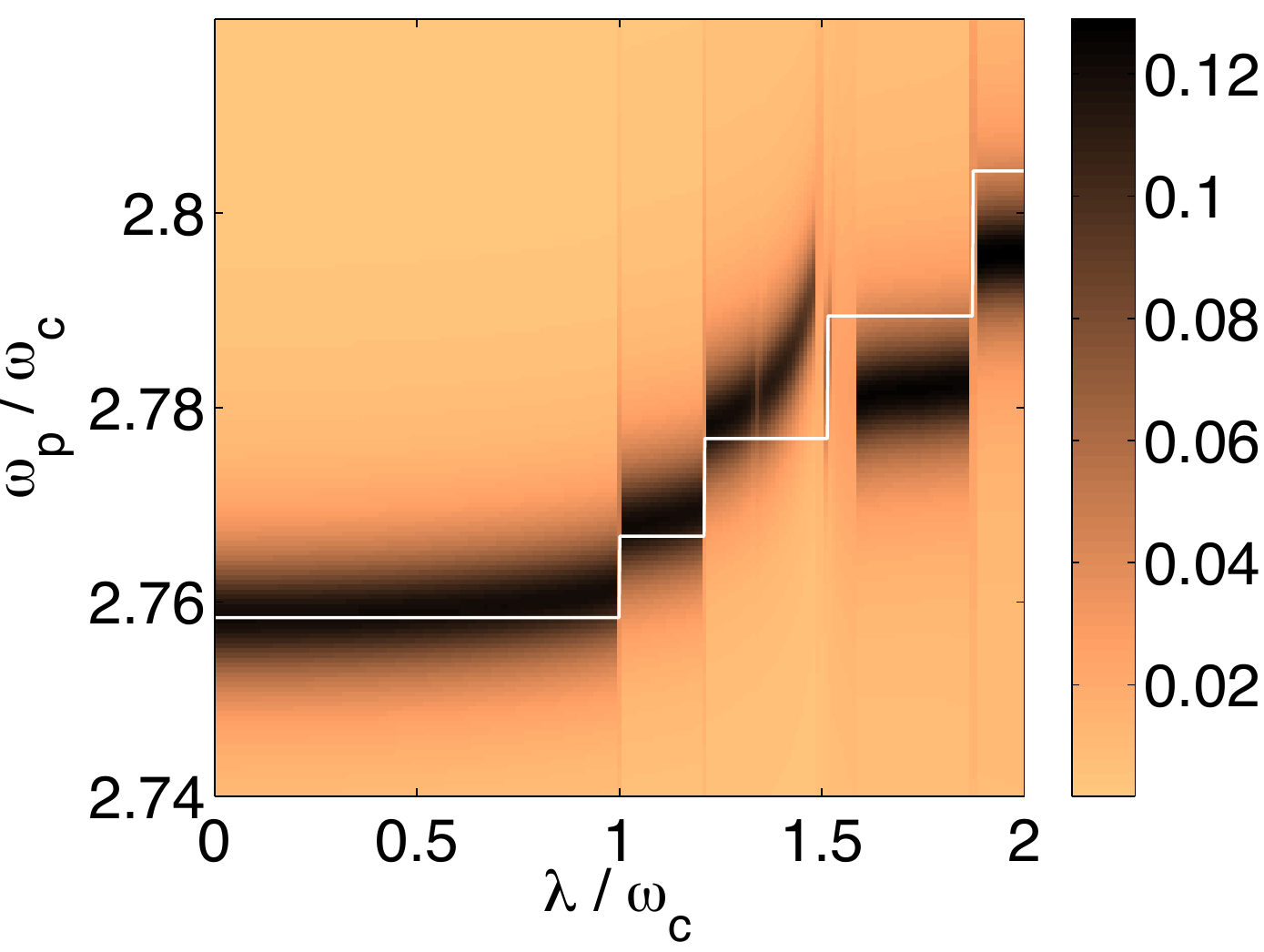}
       \end{center}
 \caption{
Excited state population of the ancilla qubit $M$ versus the coherent drive frequency $\omega_p$ for different values of collective coupling $\lambda$.
Left panel: Dicke model with $\Omega_p = 0.5 \gamma_M$. Right panel: Tavis-Cummings model with $\Omega_p = 0.2 \gamma_M$. Dissipation parameters: $\gamma_M = \gamma_c = \gamma_0 = 0.01 \omega_c$. The other parameters are as in Fig.~\ref{threemodels}. The white line corresponds to the analytical curve in Eq.~(\ref{dispersive}).
\label{spectra}}
\end{figure}

In order to include  dissipation consistently with the ultrastrong coupling regime, we need to consider a master equation for the density matrix where the quantum jumps occur between the actual eigenstates of the Hamiltonian $\mathcal{H}_{S+M}$ \cite{Beaudoin,Nissen}. We consider three decay channels, associated to the bosonic mode, the $N$ two-level systems and the measurement qubit, with dissipation rates
$\gamma_c$, $\gamma_0$ and $\gamma_M$ respectively (see Fig.~1). Namely:
\begin{equation}
\dot\rho=-i [\mathcal{H}(t),\rho]+ \frac{\gamma_c}{2} \mathcal{D}[\mathcal{U}[a^{\dagger} + a]]+ \frac{\gamma_0}{2} \mathcal{D}[\mathcal{U}[J_x]]+ \frac{\gamma_M}{2} \mathcal{D}[\mathcal{U}[\sigma_x^{(M)}]]
\label{master_equation}
\end{equation}
where
$ \mathcal{D}[\hat{A}]=\big( 2 \hat{A} \rho \hat{A}^{\dagger} - \rho \hat{A}^{\dagger} \hat{A} -\hat{A}^{\dagger} \hat{A} \rho \big)
$ and $
\mathcal{U}[\hat{A}]= \sum_{l l'}\Theta(\epsilon_{l'}-\epsilon_{l}) \langle l \vert \hat{A} \vert l' \rangle \vert l \rangle \langle l' \vert
$
defines the jump operators taking as argument the system operator involved in the coupling to the reservoir ($\Theta(\omega<0)= 0$  and $\Theta(\omega>0)= 1$).
Here we are considering reservoirs at zero temperature (they  can only absorb energy from the system $S + M$).
Note that if one uses the standard Lindblad equation with bare excitation operators as in Ref. \cite{Lambert}, the ground state of the whole system $\mathcal{H}_{S+M}$ is unstable and the reservoir excites the system even at zero temperature.

\begin{figure}[t!]
 \begin{center}
    \includegraphics[width=200pt]{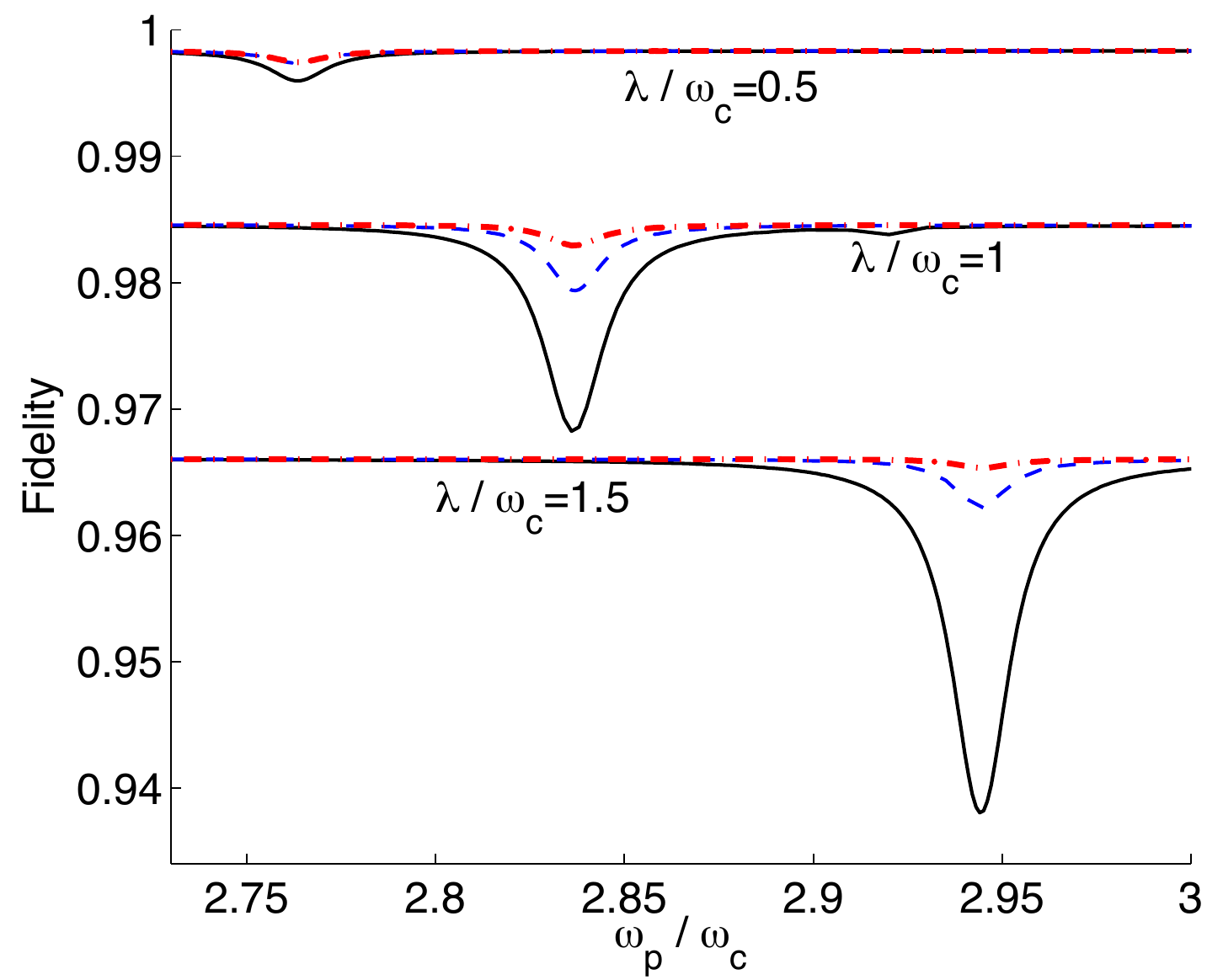}
   \end{center}
 \caption{
Measurement fidelity $\mathcal{F}$ (see definition in the text) for different values of $\lambda$ for the Dicke model
and for different dissipation rates $\gamma_c = \gamma_0 = \eta \gamma_M$ and $\gamma_M = 0.01 \omega_c$.
Solid line: $\eta = 0$. Dashed line:  $\eta = 1$. Dot-dashed line: $\eta = 10$. Other parameters as in Fig.~\ref{spectra}.
\label{fidelity}}
\end{figure}
We can now apply the master equation in Eq.~(\ref{master_equation}) to describe the spectroscopy  when qubit $M$ is driven as described by Eq.~(\ref{drivenH}). We have determined the steady-state density matrix $\hat{\rho}_{S+M}$ and consequently the reduced density matrix of qubit $M$ and system $S$ , namely $\hat{\rho}_M = Tr_{S} ( \hat{\rho}_{S+M})$ and $\hat{\rho}_S = Tr_{M} ( \hat{\rho}_{S+M})$.
In Fig.~\ref{spectra}, we show  results for the qubit excited state population $n^{(M)}_{\rm \uparrow}=  Tr_{S,M} ( \hat{\rho}_{S+M}(1 + \hat{\sigma}_z^{(M)})/2)$ versus the driving frequency $\omega_p$ and the collective atom-photon coupling $\lambda$. The ancilla excited state population spectrum shows a resonant peak that provides direct access to the vacuum-dependent qubit Lamb shift discussed so far and well described by the formula in Eq. (\ref{dispersive}).
Within our framework, we can evaluate the degree of back-action on the system $S$.  In particular we can calculate the measurement fidelity $\mathcal{F} = Tr_{S} (\hat{\rho}_S \vert G_S \rangle \langle G_S \vert)$, depending on the coupling between  $S$ and $M$, the driving of the qubit and the dissipation rates. $\mathcal{F}  = 1$ means that the cavity ground state is unaffected by the overall measurement process.
In Fig. ~\ref{fidelity}, we show $\mathcal{F}$ versus  $\omega_p$ for different values of $\lambda$ and of the dissipation rates. The moderate dip at the resonance frequency is due to creation of real excitations in the system $S$ via the driving of the qubit $M$. When the driving amplitude $\Omega_p \to 0$, the dip disappears (not shown). Out of resonance, $\mathcal{F} \to \mathcal{F}_G$, the fidelity depending only on the level mixing between qubit $M$ and system $S$ (see Fig.~\ref{N_dependence}), quantified by the ground state fidelity $ \mathcal{F}_G$.
Concerning the dissipation, our results show that when the cavity system $S$ dissipation rates $\gamma_c$, $\gamma_0$ are much smaller than the ancilla qubit dissipation rate $\gamma_M$, then the most pronounced fidelity dip is obtained (black solid-lines in Fig. ~\ref{fidelity} are for vanishing dissipation in the system $S$). Indeed, in such conditions a significant steady-state population of excited states can be created in $S$ due to the low dissipation rates, implying that the ancilla qubit cannot 'read' faithfully the ground state of the system. Instead in the opposite limit, the fidelity dip disappears ($\mathcal{F}(\omega_{dip}) \to \mathcal{F}_G$) as excited state populations in $S$ can be dissipated efficiently.

In conclusion, we have shown theoretically that the spectroscopy of an ancillary qubit coupled to a cavity (circuit) QED system is a very sensitive probe of its ground state photon properties. The spectral Lamb shift of the ancillary qubit transition is vacuum-dependent, namely depending on the  ground state photon populations and correlations. The Lamb shift behaves qualitatively in a different way for systems described by the Dicke, Tavis-Cummings and Hopfield models, whose exotic vacua are qualitatively different. By a consistent solution of the master equation to include dissipation in the ultrastrong coupling regime, we have studied the measurement fidelity by accounting for level-mixing between system and measurement qubit, driving and dissipation.  The present work demonstrates that ancillary qubit spectroscopy of cavity QED systems is a promising tool to study non-destructively the rich physics of QED vacua in the ultrastrong light-matter coupling regime.

C. C. acknowledges partial support from Institut Universitaire de France and from ERC Consolidator grant 'CORPHO'.

\end{document}